Symmetry-Protected Weyl Band Crossings in a Triangular Altermagnet

Chao-Chun Wei[1], Xiaoyin Li[1], Sophia Adams[1,2], Jacob Kjeldahl Jensen[3], Qiang Zhang[4], Jue Liu[4], Maxim Avdeev[5,6], Dinesh Kumar Yadav[7], Vikram V. Deshpande[7], Luisa Whittaker-Brooks[3], Feng Liu[1,*], Huiwen Ji[1,*]

[1]*Department of Materials Science and Engineering, University of Utah, Salt Lake City, Utah 84112, United States*

[2]*Department of Chemical and Biomolecular Engineering, Johns Hopkins University, Baltimore, Maryland 21218, United States*

[3]*Department of Chemistry, University of Utah, Salt Lake City, Utah 84112, United States*

[4]*Neutron Sciences Division, Oak Ridge National Laboratory, Oak Ridge, Tennessee 37831, United States*

[5]*Australian Nuclear Science and Technology Organisation, Kirrawee DC, New South Wales 2232, Australia*

[6]*School of Chemistry, The University of Sydney, Sydney, New South Wales 2006, Australia*

[7]*Department of Physics & Astronomy, University of Utah, Salt Lake City, Utah 84112, United States*

Corresponding: huiwen.ji@utah.edu; fliu@eng.utah.edu.

## ABSTRACT

Weyl semimetals and altermagnets represent two distinct classes of quantum materials exhibiting nontrivial topological and magnetic order, respectively. Here we report the realization of a Weyl semimetal altermagnet in $Cr_7Se_8$, combining neutron diffraction and first-principles calculations. The hexagonal system hosts a coplanar 120$^\circ$ compensated magnetic order on a triangular lattice, which breaks inversion-time-reversal and translation-time-reversal symmetries simultaneously while preserving a crystalline mirror plane. The resulting electronic structure features linearly dispersing nodal loops close to the Fermi level ($E_F$) confined to the mirror-invariant $k_z$=0 plane. Along high-symmetry directions, the crossings near $E_F$ form Dirac-like fourfold degeneracies in the absence of spin-orbit coupling (SOC); at generic momenta, these crossings split into twofold and form continuous Weyl-like nodal loops protected by mirror symmetry, which transform into a Weyl point-nodal semimetal. The momentum-dependent spin polarization exhibits an *f*-wave-like pattern characteristic of odd-parity altermagnets.

*Introduction*—Weyl semimetals[1] and altermagnets[2] represent two distinct paradigms of quantum materials with unconventional magnetic responses and potential for spintronic functionalities. Weyl semimetals arise from symmetry-protected band crossings acting as monopoles of Berry curvature, typically enabled by the breaking of either time-reversal ($T$) or inversion symmetry ($P$), and host emergent phenomena including chiral quasiparticles[3], Fermi arc surface states[4], and anomalous transport[5]. In contrast, altermagnets are magnetically-compensated systems with vanishing net magnetization in which crystal symmetries enforce momentum-dependent spin splitting, giving rise to anisotropic spin textures and unconventional magneto-transport without relying on net magnetization[6]. While initially formulated for collinear magnetic orders, this mechanism has extended to more general spin configurations[7,8], opening the route to altermagnetic behavior in a broader class of materials.

Despite their disparate symmetry origins—explicit symmetry breaking in Weyl semimetals versus symmetry-enforced spin structures in altermagnets—both classes exhibit pronounced Berry-curvature-driven responses. Theoretical proposals suggested that altermagnetic systems can host Dirac and Weyl crossings[9], but experimentally established materials simultaneously exhibiting altermagnetic spin splitting and non-trivial band topology remain scarce[10]. Here we show that $Cr_7Se_8$, a Cr-deficient derivative of the NiAs structure, is a Weyl semimetal altermagnet. Neutron diffraction and magnetic property measurements establish a compensated magnetic order featuring a 120° coplanar spin configuration with a magnetic propagation vector $\boldsymbol{k}$ = (1/3, 1/3, 0) below $T_N$, which differs from the "umbrella"-type magnetic order of stoichiometric CrSe[11] by the absence of out-of-plane moments. First-principles calculations demonstrate that the resulting magnetic symmetry generates both altermagnetic spin splitting and mirror-protected nodal loops in the vicinity of $E_F$ without SOC. These loops comprise Dirac-like crossings along high-symmetry lines and Weyl-like crossings at generic momenta, forming continuous topological nodal loops in the absence of SOC, accompanied by an $f$-wave–like alternating out-of-plane ($S_z$) spin splitting. Upon SOC is present, the Dirac crossings gap out whereas symmetry-protected Weyl-like point crossings remain.

*Structural and Magnetic properties*—$Cr_7Se_8$ adopts a hexagonal NiAs-type crystal structure with random distribution of Cr vacancies, according to the refinements of neutron diffraction at room-temperature (Figs. 1d, S1[12]) as well as single-crystal X-ray diffraction at 100 K (Tables S1-S3[12]). The structure, shown in Fig. 1a, features face-sharing $CrSe_6$ octahedra that are stacked between the layers. Two antiferromagnetic (AFM) transitions at ≈220 K and ≈50 K were identified using DC magnetization on parallel-stacked mm-sized single-crystal platelets (Fig. 1b). Notably, the 220 K AFM transition is observed only when the field is parallel to the *ab* plane, suggesting that the corresponding Néel vector lies within the plane (Fig. 1b). Meanwhile, a bifurcation is observed below the 50 K transition, which is more pronounced in the presence of an out-of-plane field. The frustration parameter obtained based on the Curie-Weiss fit of magnetic susceptibility (Fig. S3[12]), $f = |\theta_{CW}|/T_N$, is ~0.37, indicating minimal magnetic frustration yet strong magnetic exchange interactions that partially cancel each other. The $Cr_7Se_8$ phase we report here is different from the monoclinic phase in literature, which was synthesized at a much lower temperature and undergoes an AFM transition at 147 K, with weak magnetic anisotropy and a spin-flop transition at high fields[13].

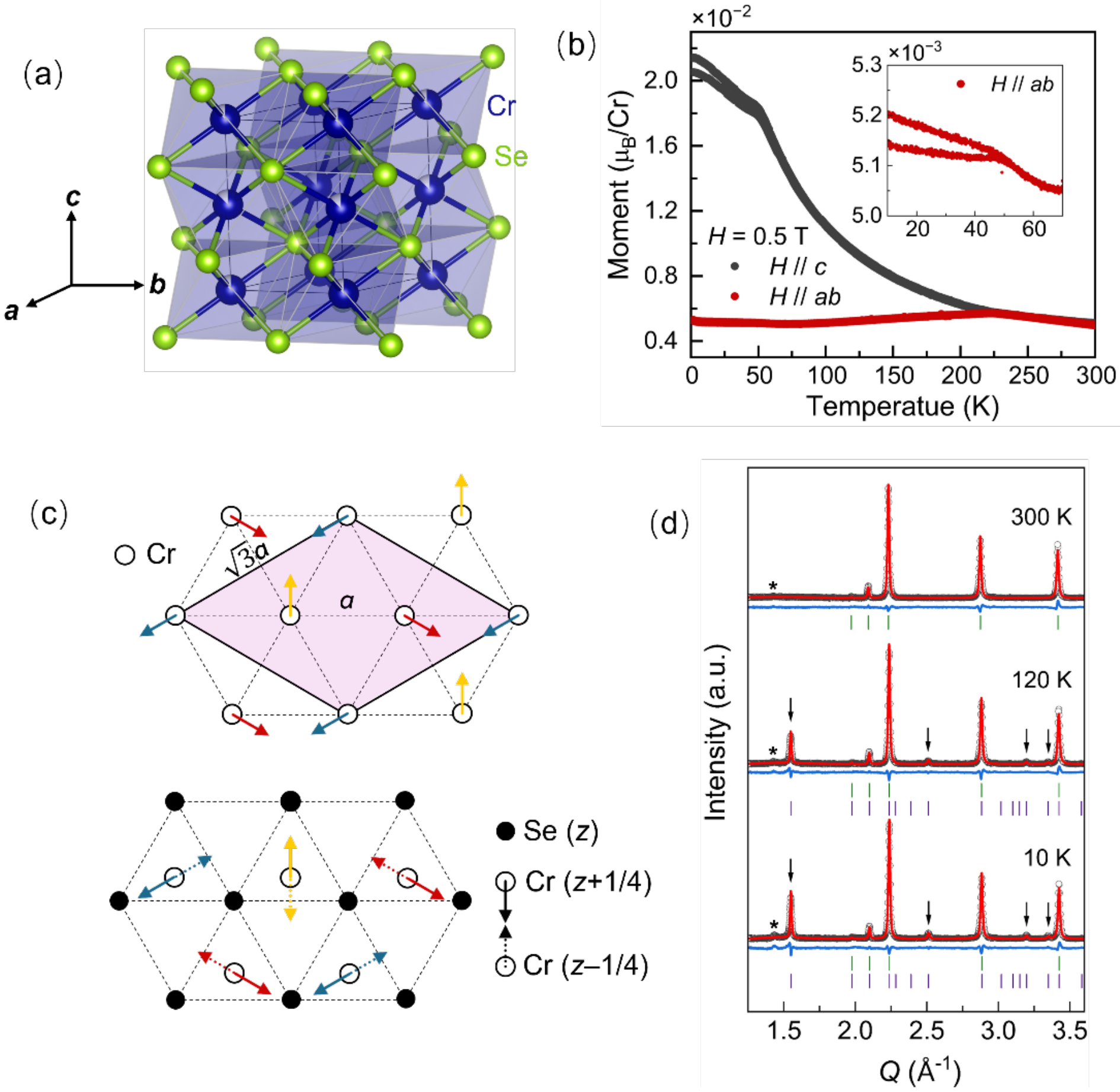

**Figure 1. Magnetic properties of $Cr_7Se_8$.** (a) Crystal structure of the NiAs type with randomly distributed Cr vacancies. (b) Magnetization (*M*) vs. temperature (*T*) on multiple parallel-stacked single crystals. To compensate for imperfect alignment, *M* measured under *H*//*c* is multiplied by a factor of 1.12 to overlap with *M* under *H*//*ab* in the paramagnetic region. Low-*T* region under *H*//*ab* is shown in the inset. (c) Coplanar magnetic structure, with a $\sqrt{3} \times \sqrt{3}$ superlattice highlighted in pink in a representative Cr layer. Open and solid dots represent Cr and Se atoms, respectively. Arrows represent Cr moments. (d) Neutron diffraction refinement. Black dots, red/blue curves, and green/purple ticks represent observation, calculated/difference and Bragg/magnetic peak positions, respectively. The small hump slightly below 1.5 $Å^{-1}$ (*) exists at 300 K and develops at low temperatures and is likely due to a minor magnetic impurity.

Neutron diffraction reveals a coplanar triangular magnetic order below the high-*T* transition. Figs. 1d and S1[12] show the diffraction patterns at 120 K and 10 K, below the two transition temperatures identified. At 120 K, additional magnetic Bragg peaks emerge (highlighted by black arrows in Fig. 1d), which are indexed by a non-zero magnetic vector $\boldsymbol{k}$ = (1/3, 1/3, 0) and magnetic space group $P\bar{6}2'm'$. The corresponding magnetic structure (Fig. 1c) hosts in-plane moments 120° rotated relative to each other and AFM interlayer alignment, yielding a fully-compensated triangular structure. The structure is consistent with the single-crystal magnetization measurements, i.e., the ordered magnetic moments lie within the *ab* plane. At 10 K, no additional magnetic peaks are observed although the existing magnetic peaks grow in intensity. Accordingly, the refined moment increases from 1.93 $\mu_B$/Cr at 120 K to 2.37 $\mu_B$/Cr at 10 K. This indicates that the low-$T$ anomaly in magnetization is not associated with a second bulk magnetic transition but due to defects. Accordingly, the bifurcation between ZFC and FC associated with the low-$T$ transition might arise from domain-wall effects or spin-glass-like behavior. The ordered Cr moment is smaller than that obtained from a Curie-Weiss fit, i.e., 4.33 $\mu_B$/Cr atom (Fig. S3[12]), likely because unpaired electrons remain partially itinerant or frustrated at 10 K.

This magnetic structure satisfies the two symmetry criteria of altermagnets or more broadly non-relativistic spin-splitting antiferromagnets[14]. First, the magnetic structure breaks the translation-time reversal ($\tau T$) symmetry. $\tau$ is a translation symmetry of the atomic structure. In our case, the non-zero magnetic vector $\boldsymbol{k}$ = (1/3, 1/3, 0) results in an expanded superlattice of $\sqrt{3}a \times \sqrt{3}b$ (Fig. 1c, upper). However, the 120$^\circ$ arrangement of Cr moments precludes an in-plane $\tau$ to be paired with $T$ without a concurrent 3-fold rotation. The $PT$ symmetry is also broken. Since opposite moments exist only between neighboring Cr layers, any inversion center that could combine with $T$ would have to lie within the Se layer (Fig. 1c, bottom). However, this is ruled out by the triangular arrangement of the Se atoms. The absence of $\tau T$ and $PT$ thus permits a momentum-dependent spin polarization, which should obey the same symmetry relations as those in real space.

*Odd-parity altermagnetism*—The anticipated non-relativistic spin splitting is explored by first-principles calculations (see the SI[12] for method details). A stoichiometric composition of CrSe with the experimentally determined coplanar magnetic structure was used. The compositional approximation is justified because non-relativistic altermagnetic spin splitting is mainly a result of crystal symmetry and magnetic order, and based on a rigid-band assumption. In the absence of SOC, the calculated band structure, shown in Fig. 2a-c, exhibits pronounced momentum-dependent spin splitting. This behavior is in stark contrast to conventional ferromagnets, where the sign of spin splitting is nearly uniform throughout the Brillouin zone. The spin polarization is dominated by the out-of-plane component $S_z$ (Fig. 2c), while the $S_x$ and $S_y$ components are strongly suppressed (Fig. 2d, e). The results suggest spin splitting through a fictitious out-of-plane exchange field despite the in-plane moments and can be understood within the spin-space-group classification recently developed for odd-parity altermagnetic systems[15]. Here, the spin texture of CrSe satisfies $s_z(k) = -s_z(-k)$ while $s_{x,y}(k) = 0$, a spin configuration that is symmetrically allowed for coplanar magnetic systems. Furthermore, the $S_z$-projected bands are spin-split except along $\Gamma - K$ equivalent directions, forming a characteristic six-lobe pattern related by 3-fold rotation symmetry (Figs. 2a, b and S4[12]). The spin polarization reverses sign under the rotation. The pattern is characteristic of an $f_{x^3-3xy^2}$ -wave-like angular dependence of odd-parity altermagnets. Notably, near $E_F$, linearly dispersed bands span over a wide energy window of nearly 1 eV. These bands are along in-plane directions and are expected to make dominant contribution to the transport properties of the material. Since the actual material $Cr_7Se_8$ is electron-deficient compared with the computed iso-structural CrSe, emergent transport properties, such as the anomalous Hall effect, are likely to be measured in an electron-doped $Cr_7Se_8$ or through ionic liquid gating.

The linear-dispersed bands near $E_F$ are contributed by Cr $d$ orbitals, as suggested by the apparent valences of 2– for Se and nearly 2+ for Cr and confirmed by the density of states (DOS) calculation in Fig. S6a[12]. Due to the $D_{3d}$ point group symmetry around Cr, its $d$ orbitals split into low-energy $a_{1g}$, $e_g(\pi)$ and high-energy $e_g(\sigma)$, which are all partially filled in a high-spin state ($4s^2 3d^4$). The $e_g$ orbitals from $d$xz, $d$yz and $d$xy, $dx^2–y^2$ dominate the linear dispersions while $a_{1g}$ from $dz^2$ has little contribution (Figs. S6b, S7[12]).

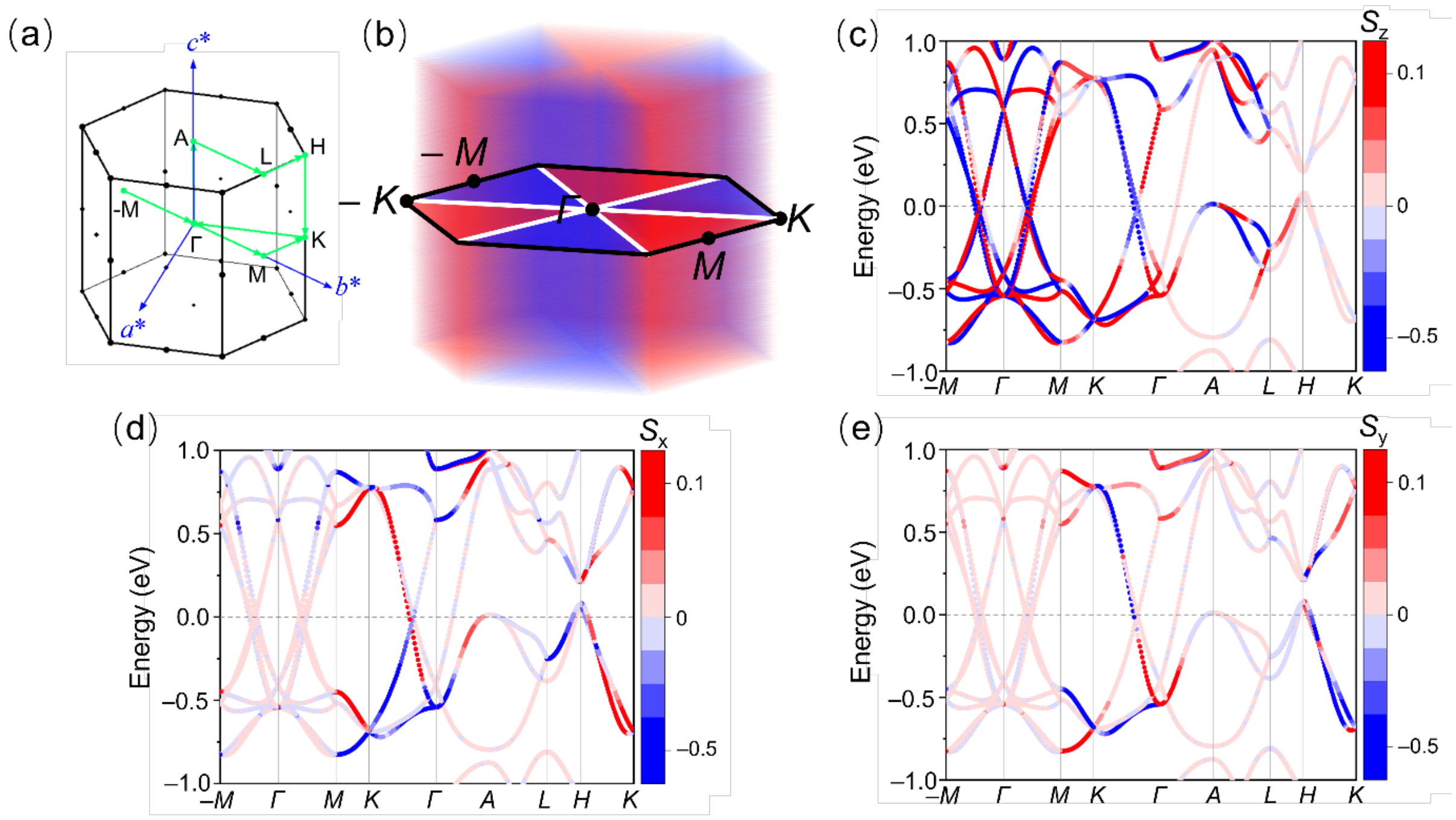

**Figure 2. Altermagnetic spin splitting in CrSe based on a triangular coplanar magnetic order.** (a) Brillouin zone showing high-symmetry paths. (b) Schematic of six-lobe *f*-wave-like spin texture in the momentum space with Zeeman nodal planes in between. (c) Calculated electronic structure projected onto the $S_z$ component. Blue and red dispersions represent spin-up and spin-down bands, respectively. (d,e) Calculated electronic structure projected onto the in-plane spin components $S_x$ and $S_y$.

*Altermagnetic Weyl band crossings*—Remarkably, the linearly dispersed spin-split bands near $E_F$ form continuous nodal loops encircling the $\Gamma$ point in the $k_z = 0$ plane without SOC. The low-energy electronic structure is mapped out in Fig. 3a. Six bands are present for these crossings. Along $K - \Gamma - K$ and symmetry-equivalent paths, the bands remain doubly degenerate; and three such doubly degenerate bands intersect to form Dirac-like fourfold-degenerate nodal points (Fig. 3b, c). Away from these paths, non-relativistic spin splitting lifts the degeneracy, resulting in Weyl-like twofold crossings (Fig. 3b, d). These Dirac- and Weyl-like features trace out continuous nodal loops (Fig. 3e) in the momentum space with a radius of approximately $0.2 \times 2\pi/a$ from $\Gamma$ ($a$ is the in-plane lattice constant of the magnetic superlattice). The spin-split bands reverse their polarization sign along the nodal loops each time they transverse an $K - \Gamma - K$ equivalent direction.

To understand the origin of the nodal loops, we compare triangular magnetic structures with and without out-of-plane moments (see Fig. S8[12]). Linearly-dispersed bands and nodal loops emerge exclusively in the coplanar AFM state, whereas magnetic configurations with finite out-of-plane moments do not exhibit such features. From a symmetry point of view, by reducing the out-of-plane moments and transitioning from a magnetic symmetry of $P6_3'/m'm'c$ for the collinear type (Fig. S8c[12]) to $P\bar{6}2'm'$ for the coplanar type, a horizontal mirror symmetry is restored. The nodal loops arise from band crossings between states with different mirror eigenvalues, which prevents hybridization in the absence of SOC. Such interplay between triangular magnetic order and crystalline symmetry thus produces a unique electronic structure combining altermagnetism and topologically non-trivial crossings. When SOC is included, the Dirac nodes along $K-\Gamma-K$ are gapped while half of the Weyl nodes along other $k_z = 0$ directions are still protected (Fig. S5[12]), and the system becomes a Weyl point-nodal semimetal.

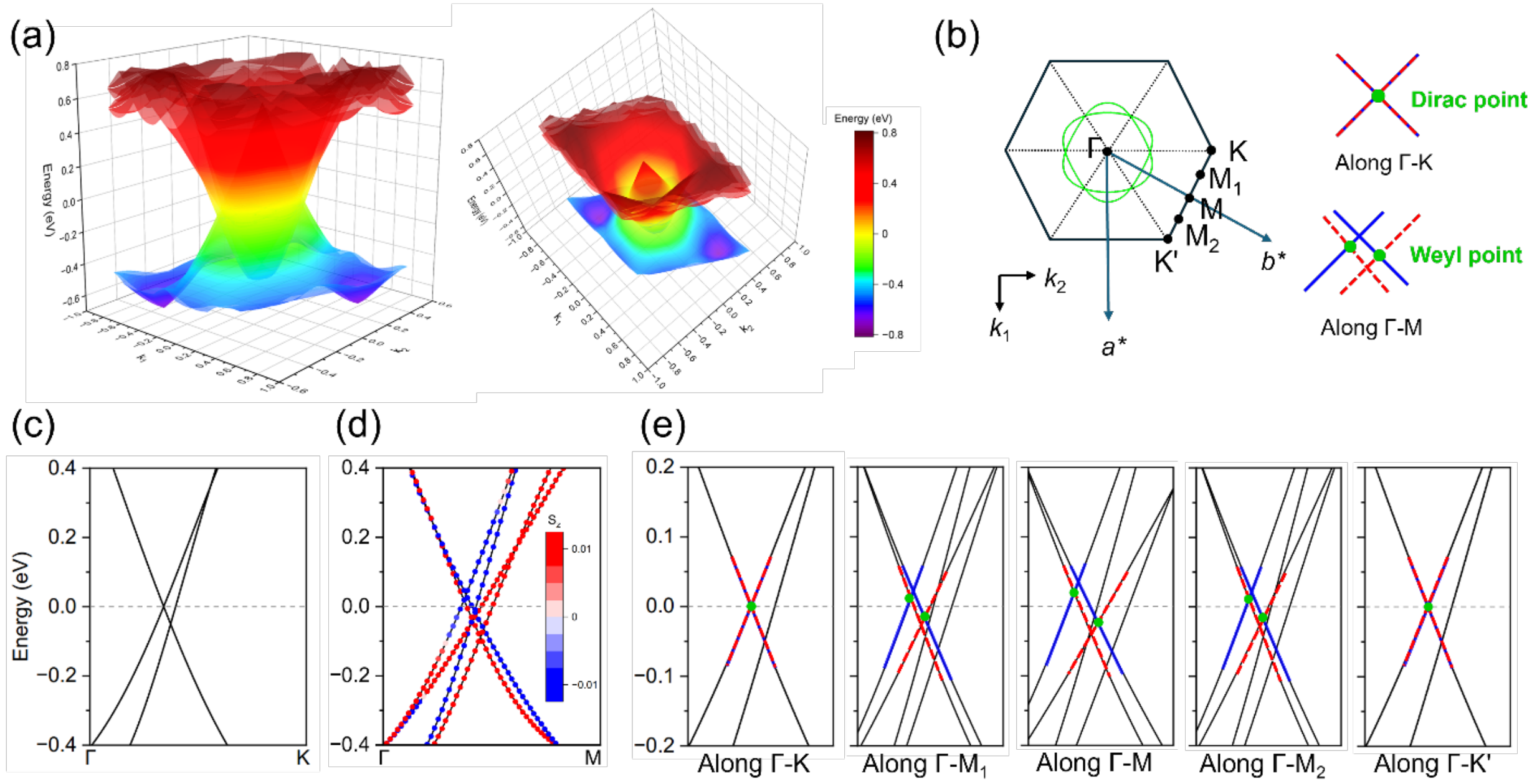

**Figure 3. DFT calculated Weyl nodal loops in CrSe without SOC.** (a) Three-dimensional band structure in the $k_z = 0$ plane showing closed nodal loops around $\Gamma$. (b) Schematic illustration of the nodal loop formed by band crossings. Four bands are used for illustrations. Along $-K-\Gamma-K$, the bands remain doubly degenerate and their crossings form Dirac-like fourfold nodal points; away from these paths, non-relativistic spin splitting lifts the degeneracy, resulting in Weyl-like twofold crossings. The nodal points near the Fermi level (marked by green dots) trace continuous loops (green lines) in momentum space. (c) Band dispersions along $\Gamma-K$ forming Dirac-like fourfold crossings. (d) Band dispersions along $\Gamma-M$ forming Weyl-like twofold crossings. (e) Band dispersions along $\Gamma-K$, $\Gamma-M_1$, $\Gamma-M$, $\Gamma-M_2$, $\Gamma-K'$ showing the evolution from Dirac to Weyl crossings along the nodal loop. Four representative bands are highlighted in color, with their crossings near $E_F$ marked by green dots.

*Discussion*—Given the critical connection between the coplanar triangular magnetic order and the Weyl band crossings, controlled defects such as Cr vacancies might provide a viable route toward engineering altermagnetic topological phases. The NiAs-type magnets are known to have rich magnetic ground states (FM, collinear in-plane or out-of-plane AFM, coplanar AFM, and their combinations) that are highly sensitive to the relative strengths of exchange interactions[16]. The 12.5% Cr vacancies in $Cr_7Se_8$ are accompanied by minimal lattice contraction ($a$ = 3.679 Å, $c$ = 6.008 Å, as refined from neutron diffraction), relative to CrSe ($a$ = 3.694 Å, $c$ = 6.061 Å)[17]. Consequently, the Cr vacancies effectively increase the average Cr–Cr separation. Indeed, we found an increasing energy gain for the coplanar order relative to the out-of-plane collinear AFM order for CrSe as the lattice stretches (Fig. S9a[12]), which likely explains the transition from the reported umbrella-type AFM order in stoichiometric CrSe[11] to the purely coplanar order in $Cr_7Se_8$. Meanwhile, our room-temperature neutron pair distribution function refinements (which characterized local atomic short-range order) indicate that locally Cr vacancies condensate in parallel columns within the *ab* plane (Fig. S10[12]). Compared to random distribution of Cr vacancies, which would evenly dampen all magnetic exchange interactions, the detected Cr-vacancy short-range order would least alleviate $J_2$ (the in-plane nearest-neighbor interaction) compared to $J_1$ or $J_3$ (which are the out-of-plane nearest and next-nearest interactions, respectively). As a result, the local magnetic order around the vacancies shifts towards a higher value along the $J_2/J_1$ axis, thus re-entering the A-type AFM region (Fig. S9b[12]). This likely explains the low-$T$ magnetic anomaly near 50 K as the emergence of vacancy-induced local AFM order with moments oriented along the out-of-plane direction, rather than bulk magnetic ordering. The Cr-Se series, with particularly diverse Cr-deficient compositions and vacancy orders[13,18-22], thereby provides a fertile platform to explore the relationships between magnetic order and electronic properties.

The coexistence of altermagnetic spin splitting and Weyl band crossings in $Cr_7Se_8$ represents a new class of quantum materials. Unlike conventional Weyl semimetals, the nodal states here are embedded in a spin-split background with $\boldsymbol{k}$-dependent spin polarization, which are detectable by angle-resolved photo-emission spectroscopy. The combination is also expected to give rise to unconventional transport phenomena, such as the anisotropic anomalous Hall effect and spin-polarized currents without net magnetization. The large energy scale of the nodal states further enhances their experimental relevance. Since our DFT computation was done on the stoichiometry of CrSe, assuming rigid bands, the presence of Cr vacancies is expected to shift $E_F$ below the Weyl nodal points. This mismatch can be addressed by isoelectronic doping in CrSe with non-magnetic atoms especially when a bulkier dopant can exert chemical strain in the lattice to favor the coplanar magnetic order. Alternatively, external strain is also applicable.

**Acknowledgement**

H.J. and C.-C.W. acknowledge the support from an NSF Career grant No. 2145832. X.L. and F.L. acknowledge the support from the DOE-BES (No. DE-FG02-04ER46148). A portion of this research used resources at the Spallation Neutron Source, a DOE Office of Science User Facility operated by the Oak Ridge National Laboratory. The beam time was allocated to POWGEN on proposal number IPTS-35963.1. We have used computational resources from CHPC of the University of Utah and the DOE-NERSC. Some of the experimental data were acquired using a Physical Property Measurement System funded by the NSF under award number DMR 2319964.

# Supplemental Material

## Symmetry-Protected Weyl Nodal Loops in a Triangular Altermagnet

Chao-Chun Wei[1], Xiaoyin Li[1], Sophia Adams[1,2], Jacob Kjeldahl Jensen[3], Qiang Zhang[4], Jue Liu[4], Maxim Avdeev[5,6], Dinesh Kumar Yadav[7], Vikram V. Deshpande[7], Luisa Whittaker-Brooks[3], Feng Liu[1,*], Huiwen Ji[1,*]

[1]*Department of Materials Science and Engineering, University of Utah, Salt Lake City, Utah 84112, United States*

[2]*Department of Chemical and Biomolecular Engineering, Johns Hopkins University, Baltimore, Maryland 21218, United States*

[3]*Department of Chemistry, University of Utah, Salt Lake City, Utah 84112, United States*

[4]*Neutron Sciences Division, Oak Ridge National Laboratory, Oak Ridge, Tennessee 37831, United States*

[5]*Australian Nuclear Science and Technology Organisation, Kirrawee DC, New South Wales 2232, Australia*

[6]*School of Chemistry, The University of Sydney, Sydney, New South Wales 2006, Australia*

[7]*Department of Physics & Astronomy, University of Utah, Salt Lake City, Utah 84112, United States*

**Email*: huiwen.ji@utah.edu; fliu@eng.utah.edu.

## A. Methods for synthesis, structural and magnetic characterizations, and first-principles calculations

$Cr_7Se_8$ powder was synthesized by mixing Cr powder (Sigma Aldrich, 99%) and Se shots (Thermo Scientific, 99.999%) in a 7:8 atomic ratio. The precursors were ground, pelletized, and sealed in a quartz ampule. The sealed sample was first heated to 650°C for 48 h, followed by regrinding, and a second heat treatment at 950°C for 48 h, followed by air quenching. The pelletized $Cr_7Se_8$ powder was then sealed under vacuum with Sn shots (Luciteria, 99.99%) in a 1:5 weight ratio for crystal growth. The mixture was held at 950°C for 14 d, followed by centrifugation to remove Sn.

Magnetization was measured using a Quantum Design Physical Properties Measurement System (PPMS) with a vibrating-sample magnetometer (VSM) option. Susceptibility was obtained by normalizing the measured magnetization by the field. Neutron diffraction was performed on $Cr_7Se_8$ powder at POWGEN and NOMAD at SNS, ORNL. Approximately 1.3 g powder was loaded into a 6 mm diameter vanadium PAC can, which was subsequently sealed with helium exchange gas to ensure optimal thermal conductivity. Data were collected at 10, 120, and 300 K using a single neutron frame centered at a wavelength of 1.5 Å, providing a d-spacing coverage of 0.5–12 Å. All the data were reduced by subtracting the signal from the empty PAC container, with normalization performed using the difference between the vanadium standard and the empty instrument background. Neutron data was refined using the GSAS-II package[1].

Single-crystal X-ray diffraction was carried out on a Rigaku XtaLAB Synergy-R diffractometer with Mo $K\alpha$ radiation ($\lambda$ = 0.71073 Å). The crystal structure was refined using the Shelxle software. Crystal structures were visualized using VESTA.

First-principles calculations were performed using density-functional theory (DFT) as implemented in the Vienna Ab initio Simulation Package (VASP)[2]. The projector-augmented-wave method[3] and the Perdew-Burke-Ernzerhof generalized gradient approximation[4] were used to treat the electron-ion interaction and exchange-correlation functional, respectively. A plane-wave cutoff energy of 500 eV and a $\Gamma$-centered 12 × 12 × 8 $k$-mesh were employed. The total energy convergence threshold was set to $10^{-6}$ eV. The lattice parameters and atomic positions of CrSe were taken from structural solutions based on room-temperature neutron diffraction. The constrained local moment method[5] was used to fix the directions of the atomic magnetic moments and thereby obtain the experimentally determined coplanar antiferromagnetic configuration. Post-processing of the calculation results was performed using VASPKIT[6].

**B. Neutron diffraction and magnetization measurements on powder $Cr_7Se_8$.**

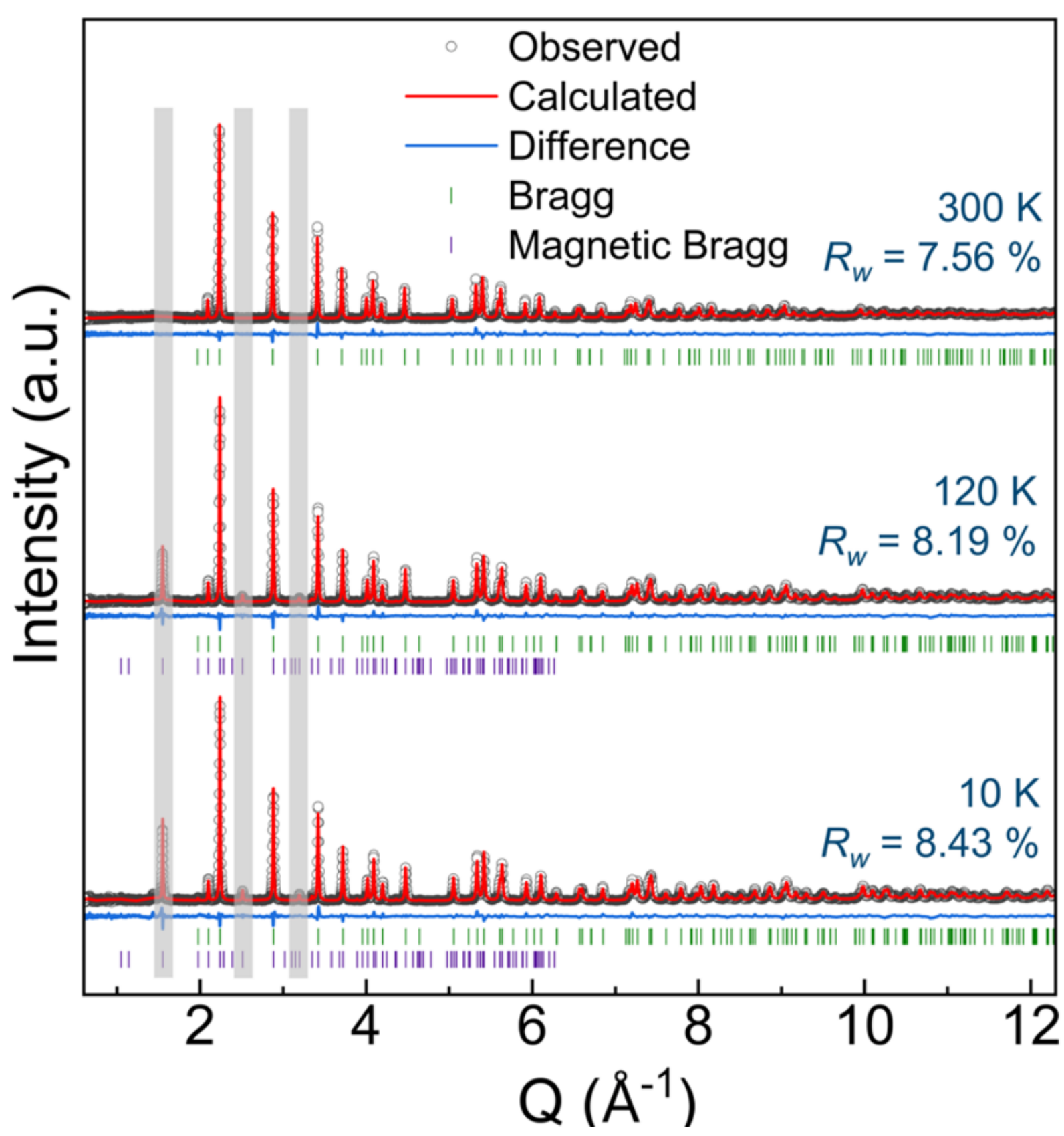

**Figure S1.** Rietveld refinements of TOF neutron diffraction patterns of $Cr_7Se_8$ collected at 300 K, 120 K, and 3 K at POWGEN. Black dots, red/blue curves, and green/purple ticks are experimental observation, calculated/difference and Bragg/magnetic peak positions, respectively. Pronounced magnetic peak positions are highlighted by grey bars.

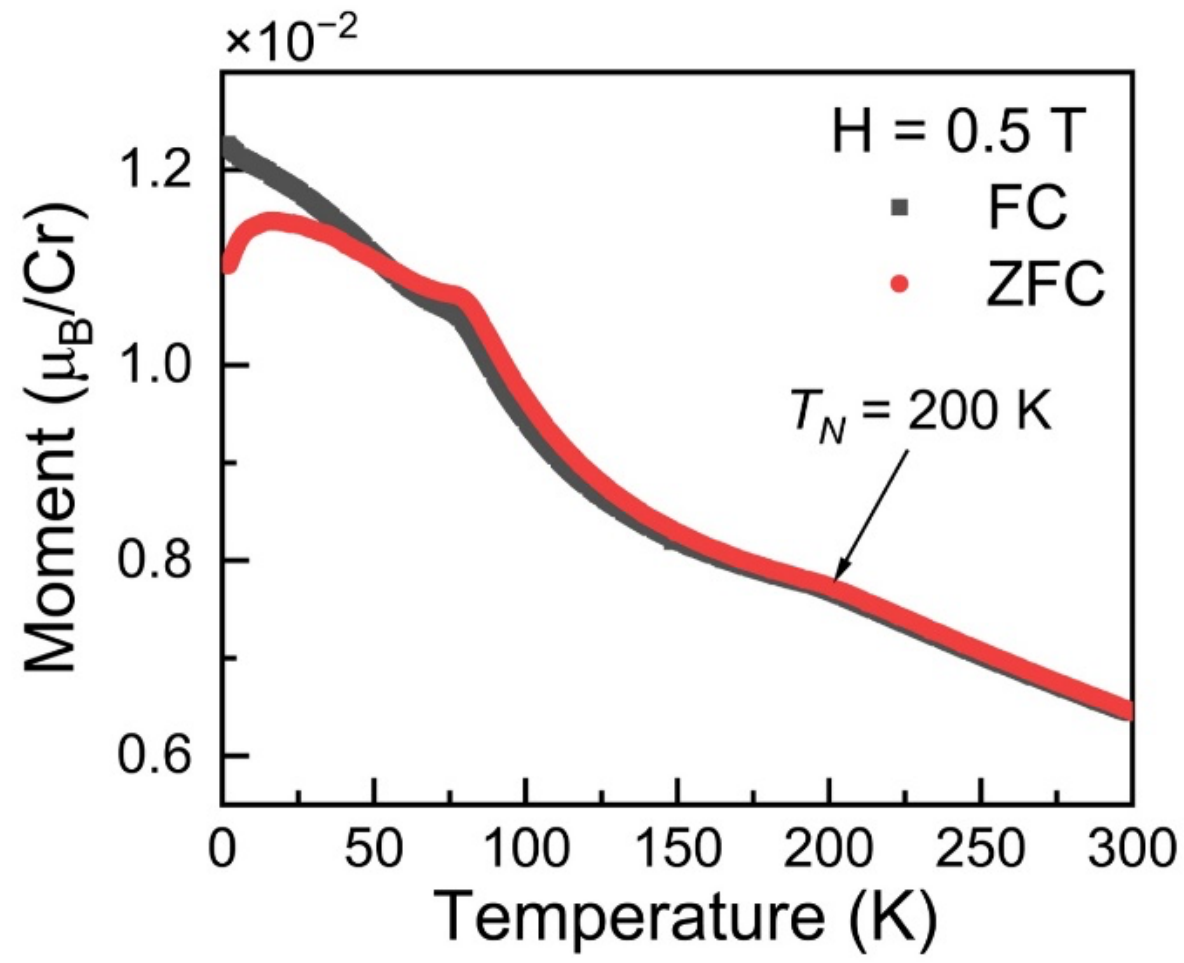

**Figure S2.** Temperature-dependent magnetization of $Cr_7Se_8$ powder under field-cooled (FC) and zero-field-cooled (ZFC) conditions.

## C. Curie-Weiss Analysis of High-Temperature Magnetic Susceptibility

Field-cooled magnetization $M(T)$ of $Cr_7Se_8$ was measured under an applied magnetic field of $H$ = 5000 Oe. The raw magnetization (emu) was converted to molar magnetic susceptibility per Cr:

$$\chi(T) = \frac{M(T)}{H} \cdot \frac{M_{molar}}{m_{sample}} \cdot \frac{1}{n_{Cr}}$$

where $M_{molar} = 995.74\ g\ mol^{-1}$ is the molar mass of $Cr_7Se_8$, $m_{sample}$ is the sample mass, and $n_{Cr} = 7$ is the number of Cr atoms per formula unit. The resulting susceptibility has units of $emu \cdot mol_{Cr}^{-1} \cdot Oe^{-1}$.

In the paramagnetic regime, the susceptibility was modeled using the Curie-Weiss law:

$$\chi(T) = \chi_0 + \frac{C}{T - \theta_{CW}}$$

where $\chi_0$ is the temperature-independent susceptibility, $C$ is the Curie constant, and $\theta_{CW}$ is the Curie-Weiss temperature.

The data in the temperature range 300-390 K were fitted using the *curve_fit* function in SciPy Python.[7] For visualization, the inverse susceptibility $\chi^{-1}(T)$ and the inverted fitted function were plotted, as shown in Figure S3.

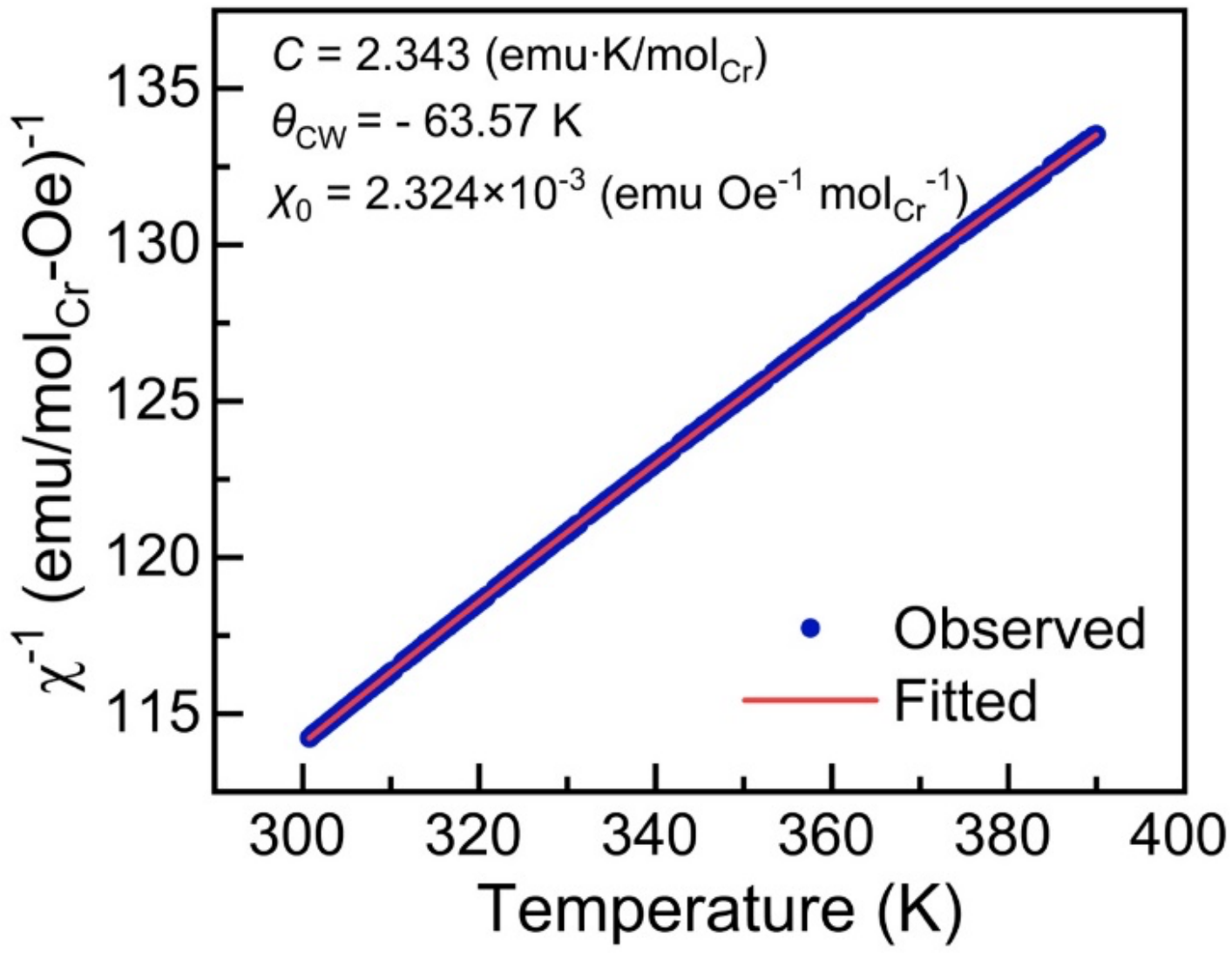

**Figure S3.** Inverse magnetic susceptibility ($\chi^{-1}$) as a function of temperature for a $Cr_7Se_8$ sample measured in the paramagnetic regime.

The fitting yields:

$$\chi_0 = (2.3237 \pm 0.0069) \times 10^{-3} emu \cdot mol_{Cr}^{-1} \cdot Oe^{-1},$$

$$C = 2.3430 \pm 0.0056\ emu \cdot K \cdot mol_{Cr}^{-1},$$

$$\theta_{CW} = -63.57 \pm 0.49\ K.$$

Note that, in cgs units, $emu \cdot Oe^{-1}$ is equivalent to $emu$ and as the result the Oe term is commonly omitted from susceptibility units such as Curie constant.[8] Also, the effective magnetic moment per Cr is obtained from the Curie constant via $\mu_{eff} = 2.828\sqrt{C}$, which yields $\mu_{eff} = 4.33\mu_B/Cr$. The negative Curie-Weiss temperature indicates dominant antiferromagnetic interactions, while the extracted effective moment is close to the expected value for mixed-valence $Cr^{2+}/Cr^{3+}$ configurations in $Cr_7Se_8$.[9]

**D. Symmetry of the non-relativistic spin splitting and the effect of SOC.**

The symmetry of the momentum-dependent spin polarization is revealed in Fig. S4, which shows the calculated electronic band structures along high-symmetry directions, considering only the dominant $S_z$ component. Figure S4a-c shows the momentum-dependent spin splitting in the $k_z = 0$ plane, with the selected $\boldsymbol{k}$ paths shown in Fig. S4a and the band dispersions plotted in Fig. S4b, c. In Fig. S4b, c, the spin polarization reverses sign under $k \rightarrow -k$ or when rotated by $60^\circ$ from $\Gamma - M$ to $\Gamma - N$. The spin polarization along these high-symmetry paths reveals a six-lobe pattern in the $k_z = 0$ plane (shown on the bottom in Fig. S4a). Accordingly, there are three spin-degenerate Zeeman nodal planes located at $k_x = k_y$, $k_x = -2k_y$, $-2k_x = k_y$ ($k_x$, $k_y$, $k_z$ correspond to $a^*$, $b^*$, $c^*$ in Fig. S4). Fig. S4d-g shows the momentum-dependent band structures in the $\Gamma MAL$ plane, along the more generic $-M' - \Gamma' - M'$, $-D_1 - P_1 - D_1$, and $-D_2 - P_2 - D_2$ directions. The reversed sign of spin splitting between $-M' - \Gamma'$ and $\Gamma' - M'$, similar to Fig. S4b, indicates that the six-lobe pattern retains and extends out of the $k_z = 0$ plane. Meanwhile, the symmetrical spin polarization along $-D_1 - P_1 - D_1$ and $-D_2 - P_2 - D_2$ (two directions that project into the opposite lobes on the $k_z = 0$ plane) indicates the absence of any additional nodal planes beyond the three already identified. These observations together support the classification of odd-parity, $f_{x^3-3xy^2}$-wave-like spin texture in CrSe.

To elucidate the evolution of the nodal states, we consider the symmetry of the coplanar triangular magnetic order described by the magnetic space group ($P\bar{6}2'm'$). In the absence of SOC, a horizontal mirror symmetry ($M_z$) leaves the ($k_z$=0) plane invariant. The nodal loops arise from crossings between bands belonging to distinct mirror sectors, which prevents their hybridization throughout the mirror-invariant plane. Along $\Gamma{-}K$ and their symmetry-equivalent directions, additional crystalline symmetries enforce band degeneracies, giving rise to Dirac-like fourfold crossings embedded within the nodal loops.

The coplanar magnetic structure further preserves three symmetry-related vertical mirror planes, each containing one of the magnetic moments and intersecting the Brillouin zone along a ($\Gamma{-}M$)-equivalent direction. With SOC, the spin and orbital degrees of freedom become entangled, lifting the symmetry protection responsible for the Dirac-like crossings along ($\Gamma{-}K$), which consequently acquire finite gaps (Fig. S5). In contrast, some Weyl crossings located on the mirror-invariant ($\Gamma{-}M$) directions remain gapless (Fig. S5). This selective survival strongly indicates that the crossing bands belong to distinct mirror sectors of the corresponding little group and therefore cannot hybridize. As a result, the mirror-protected nodal loops present in the non-relativistic limit evolve into isolated symmetry-protected point crossings along ($\Gamma{-}M$) directions near $E_F$ with SOC.

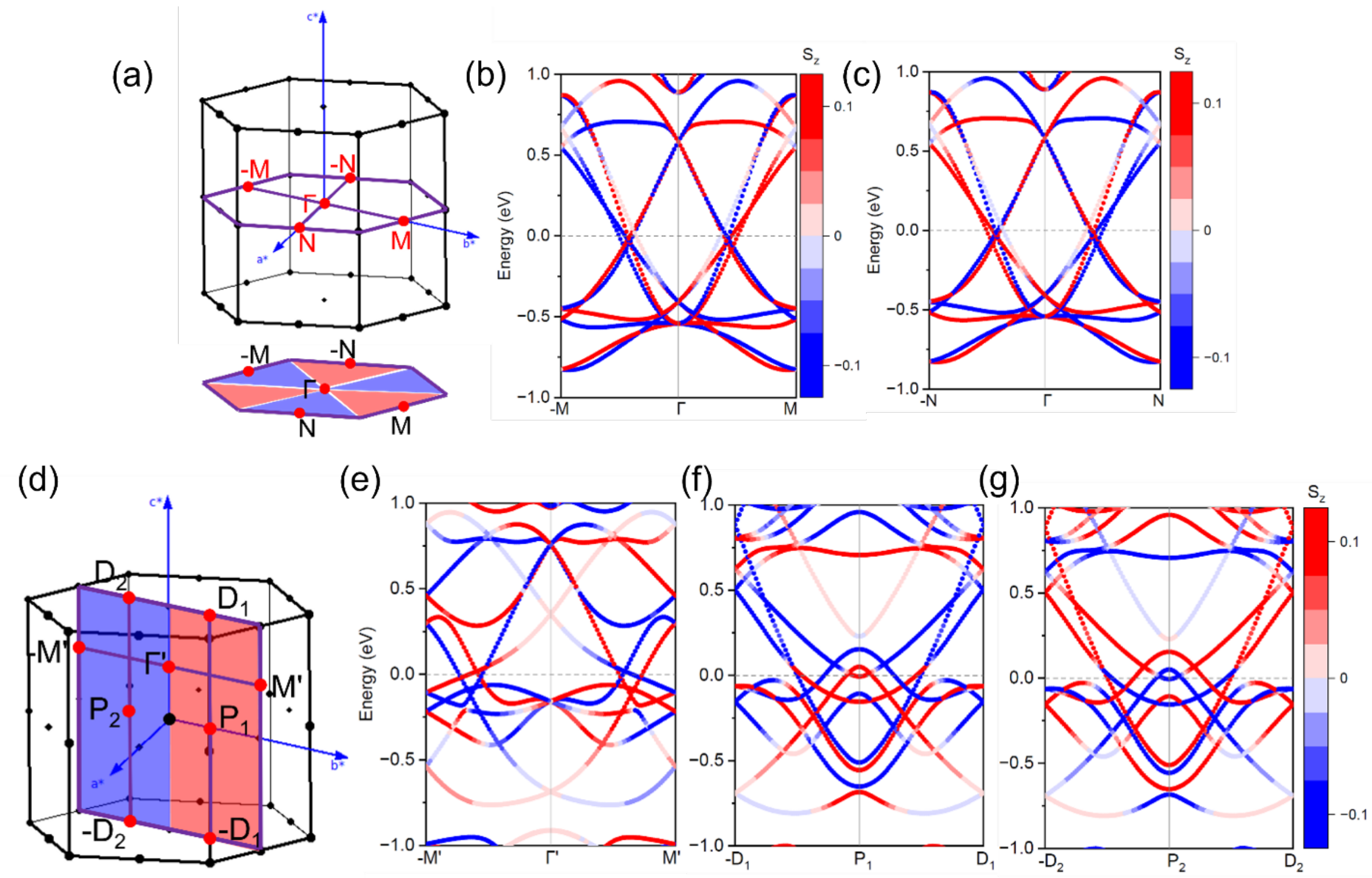

**Figure S4. Symmetry of spin splitting.** (a) First Brillouin zone with high-symmetry paths on the $k_z = 0$ plane. $S_z$-projected electronic band structures along (b) $-M - \Gamma - M$ and (c) $-N - \Gamma - N$. (d) First Brillouin zone with selected paths on the $\Gamma MAL$ plane. $S_z$-projected band structures along (e) $-M' - \Gamma' - M'$, (f) $-D_1 - P_1 - D_1$, and (g) $-D_2 - P_2 - D_2$. The blue/red color scale represents the out-of-plane spin polarization $S_z$.

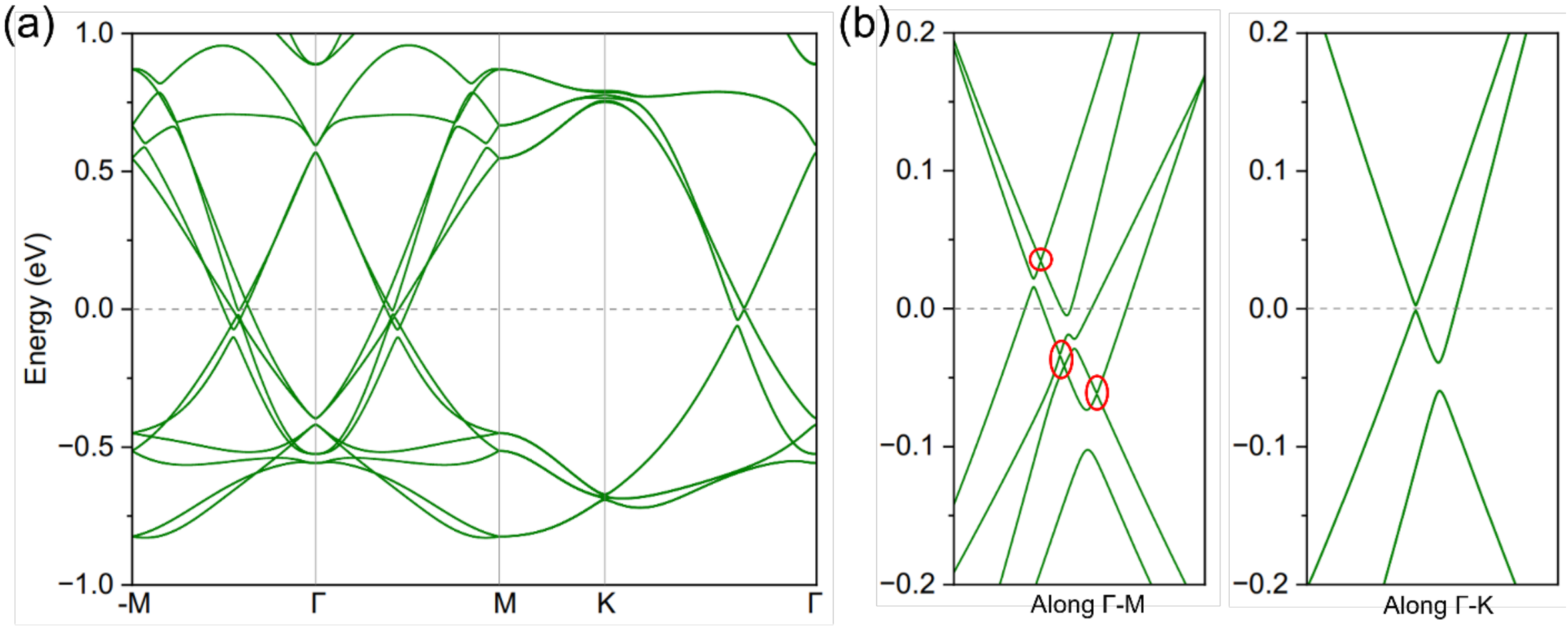

**Figure S5. Electronic band structure of CrSe with SOC.** (a) Band dispersions along high-symmetry paths. (b) Zoom-in band dispersions along Γ–M and Γ–K. Along Γ–M, half of the Weyl crossings remain gapless (highlighted by red circles). Along Γ–K, all Dirac crossings are gapped.

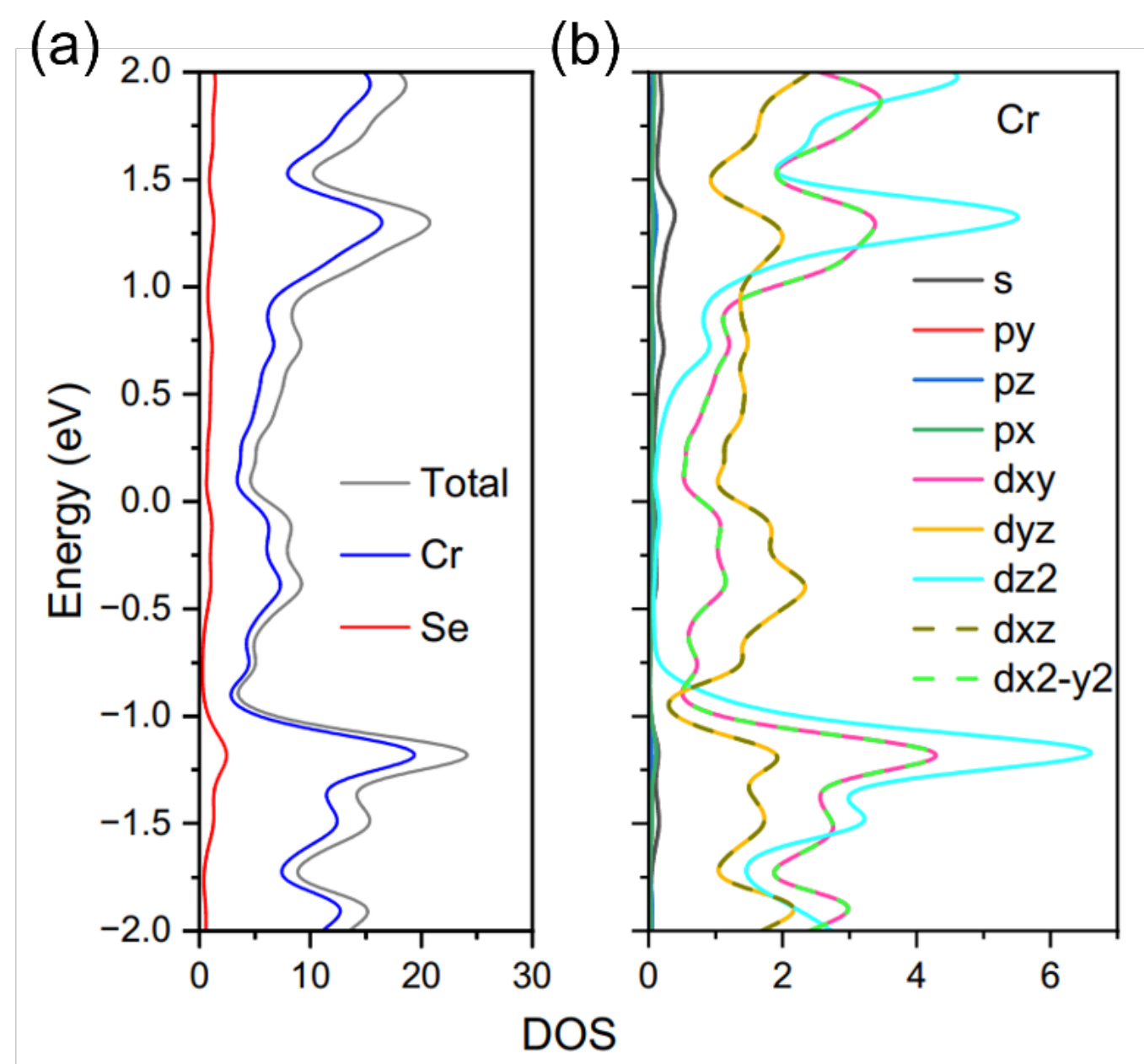

**Figure S6. Density of states (DOS) of CrSe.** (a) Total and element-projected DOS for Cr and Se. (b) Orbital-projected DOS of Cr.

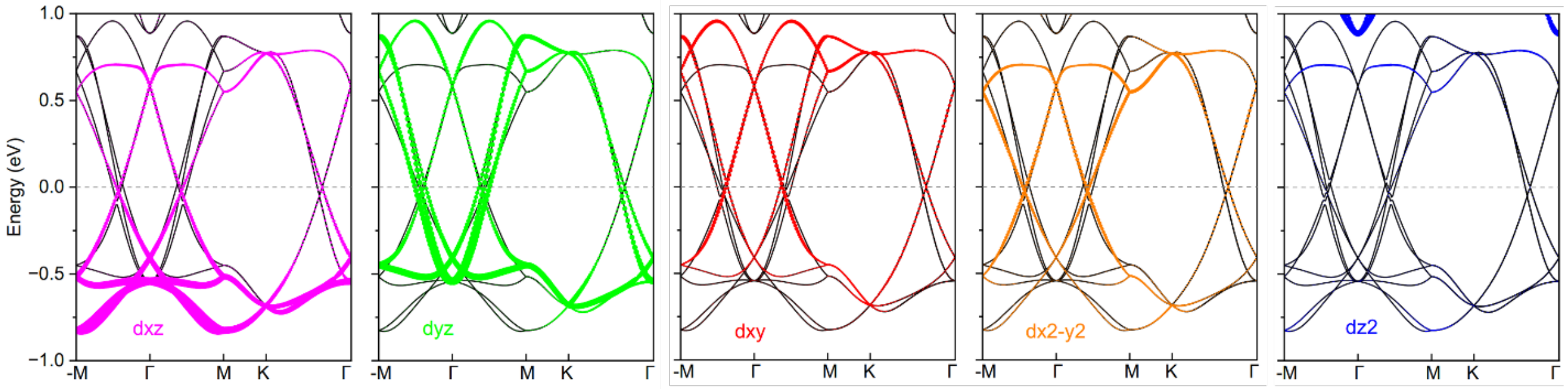

**Figure S7. Cr *d*-orbital-projected electronic band structure.** The size of the colored markers reflects the respective orbital weight. (Left to right) Purple, green, red, orange, and blue correspond to the *d*xz, *d*yz, *d*xy, $dx^2$-$y^2$ and $dz^2$ orbitals, respectively.

## E. Triangular magnetic orders with and without out-of-plane moments.

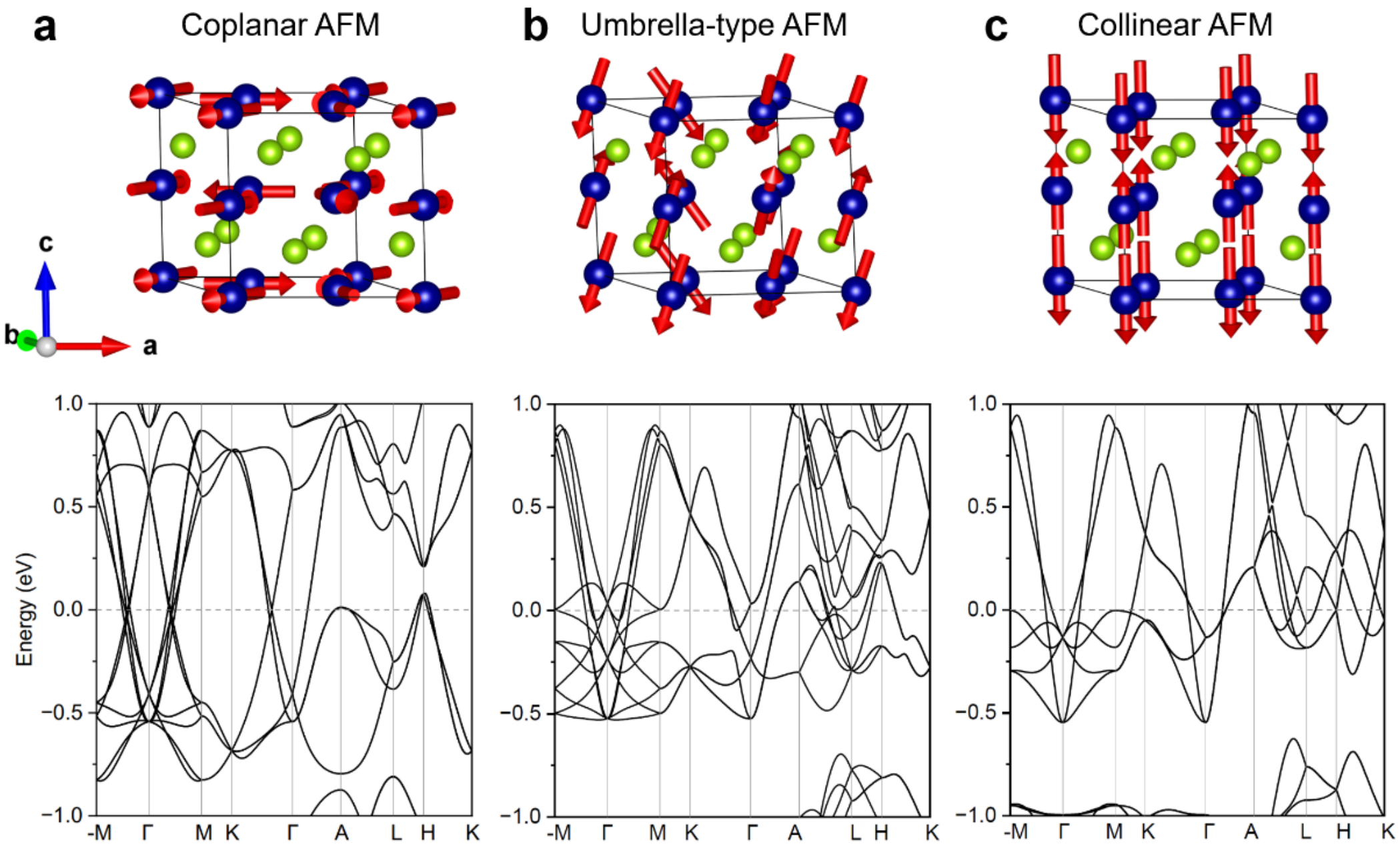

**Figure S8. Electronic band structure of CrSe with various magnetic configurations.** Magnetic structure (top panels) and their corresponding electronic bands (bottom panels) of (a) coplanar, (b) umbrella-type and (c) collinear antiferromagnetic states.

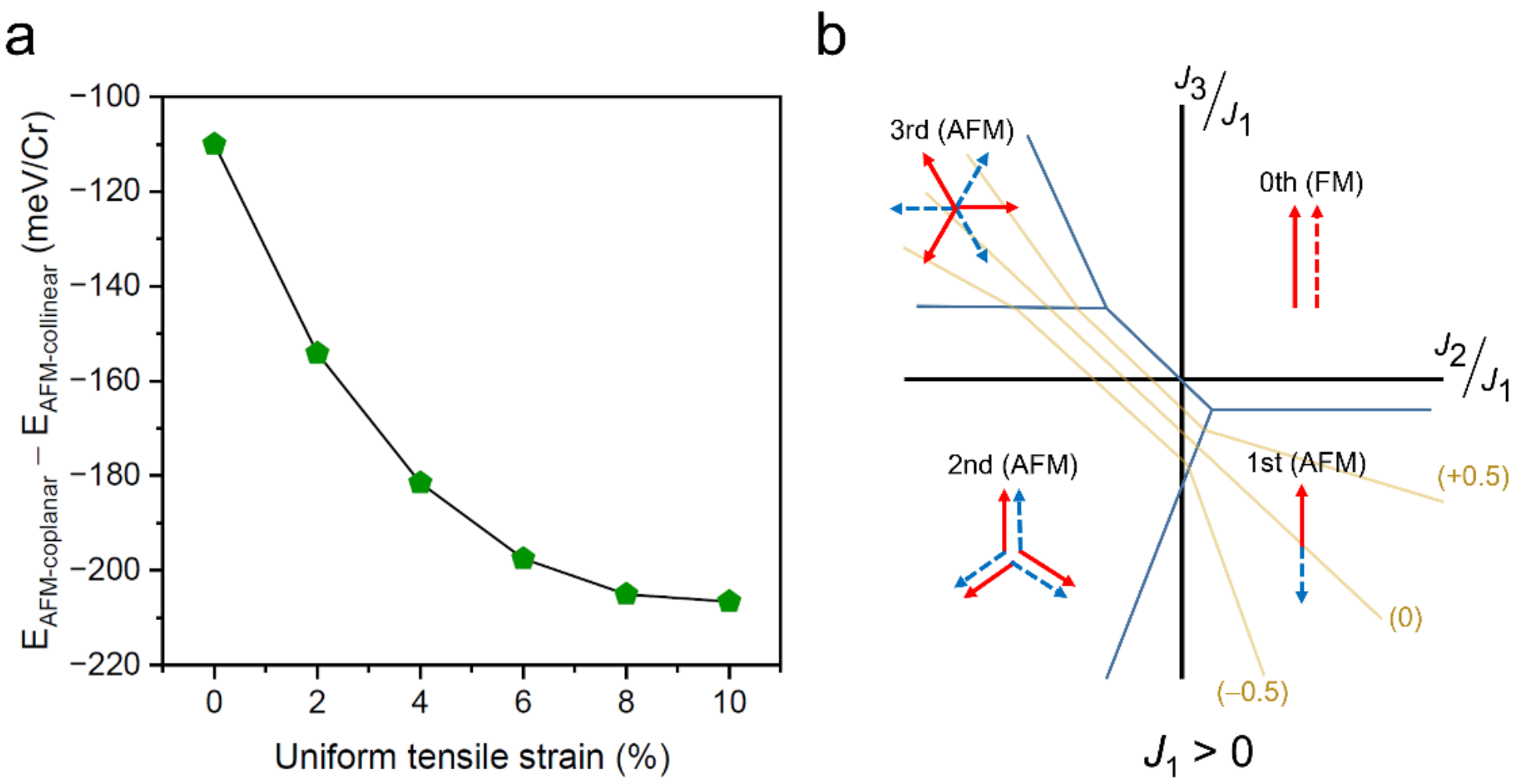

**Figure S9.** (a) DFT calculation on the energy difference between the coplanar and collinear AFM configurations (as illustrated in Figs. S8a, c) as a function of uniform tensile strain. (b) Magnetic phase diagram as a function of the exchange interaction rations $J_2/J_1$ and $J_3/J_1$ (with $J_1 > 0$) as mapped out by Hirone and Adachi, reproduced from ref [10]. $J_1$, $J_2$, and $J_3$ represent out-of-plane nearest, in-plane nearest, and out-of-plane next-nearest interactions, respectively. The yellow lines represent the boundaries with the corresponding frustration parameter $f = |\theta_{CW}|/T_N$ of +0.5, 0, and –0.5. In our case, $f$ is ~0.37

## F. Neutron PDF for solving the local Cr-vacancy order in $Cr_7Se_8$.

We performed neutron total scattering and pair-distribution function (PDF) analysis on the powder $Cr_7Se_8$ sample. The NPDF data at 300 K was refined using several distinct structural models. Their corresponding fits are summarized in Fig. S10a-d. These models are: (i) a NiAs-type structure with random distribution of Cr vacancies (Fig. S10a), which is the average model we used for neutron powder diffraction refinement; (ii) a monoclinic $Cr_7Se_8$ structure (Fig. 10b), in which Cr vacancies condensate into alternating filled and a Kagome pattern in every other layer[11]; the structure type with an equivalent but non-conventional $F2/m$ setting was reported in ref[12] for $Cr_5Se_8$. (iii) a $Cr_6Se_8$-type structure, which is reported in ref[13] ($I2/m$, Fig. S10c) that segregates Cr vacancies in alternate strips in every other layer (we assigned half occupancy in the vacant sites to bring the total Cr content to stoichiometric); (iv) finally, a $P2/m$ monoclinic structure (Fig. S10d) that we modified based on an $I2/m$ structure in (iii). Among the four models described, the $Cr_6Se_8$-derived $P2/m$ monoclinic structure (iv) provides the best fit.

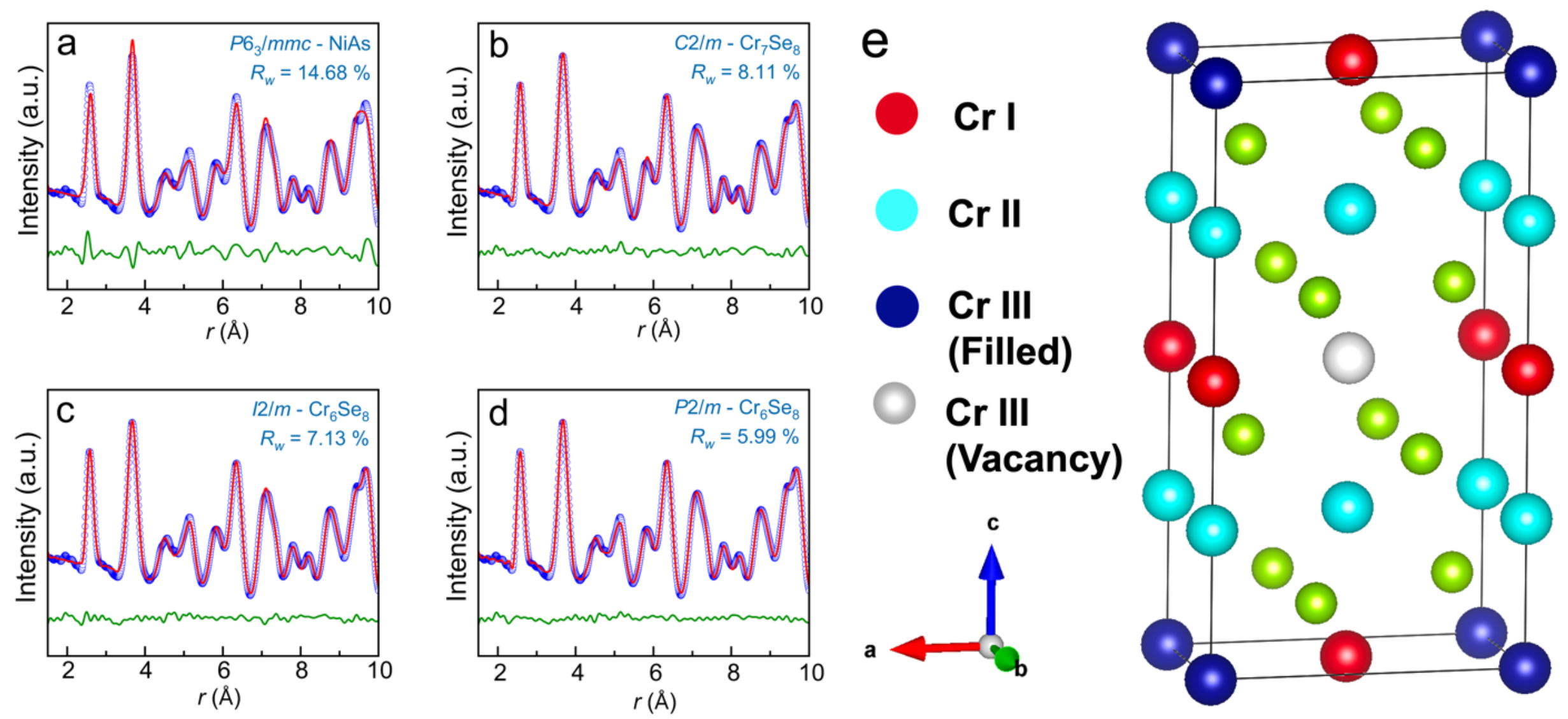

**Figure S10. Local Cr-vacancy order based on powder $Cr_7Se_8$.** NPDF refinement at 300 K using (a) a NiAs-type structure, (b) a monoclinic $C2/m$ structure model previously reported for $Cr_7Se_8$, (c) a monoclinic $I2/m$ structure derived from a reported $Cr_6Se_8$ structure, and (d) a monoclinic $P2/m$ model derived from $Cr_6Se_8$ but with vacancies ordered within columns. Blue circles, red and green curves represent observation, calculation, and fit residual, respectively. (e) Structural model based on the best fit in (d).

The transformation from the parent NiAs-type hexagonal cell (Fig. S10a) to the monoclinic $Cr_6Se_8$-type supercell (Fig. S10c, d) is given by:

$$(a_{mono}, b_{mono}, c_{mono}) = (a_{hex}, b_{hex,} c_{hex}) \begin{pmatrix} 2 & 0 & 0 \\ 1 & 1 & 0 \\ 0 & 0 & 2 \end{pmatrix}$$

The resulting $I2$/m supercell contains three crystallographically distinct Cr sites: four Type-I sites, two Type-II sites, and two Type-III Cr sites (Fig. S10e). For $Cr_6Se_8$, Type-I and Type-II Cr sites are fully occupied while Type-III sites are vacant. Previous work showed that for $Cr_{3+x}Se_4$, excess Cr tends to occupy Type-III sites.[13] Motivated by the observed site-occupancy preferences, in Fig. S10c, we adapted the structure by setting the two Type-III Cr sites to half occupancy, besides the full occupancy in the Type-I and Type-II sites, yielding a total stoichiometry of $Cr_7Se_8$. This model yields an improved fit compared to the random model ($wR$ = 14.68%, Fig. S10a). Finally, in Fig. S10d, we lifted the degeneracy of the two Type-III sites by fully occupying one site while leaving the other vacant, thereby lowering the symmetry from $I2/m$ to $P2/m$ with a refined $\beta$ close to 90° (structure shown in Fig. S10e). The final model yields a refinement with the best agreement ($wR$ = 5.99%) among all models and best captures the local Cr-vacancy order.

**G. Temperature-dependent in-plane resistivity** on a $Cr_7Se_8$ single crystal, showing metallic behavior. A kink is observed at ~210 K, close to the measured $T_N$ of the material.

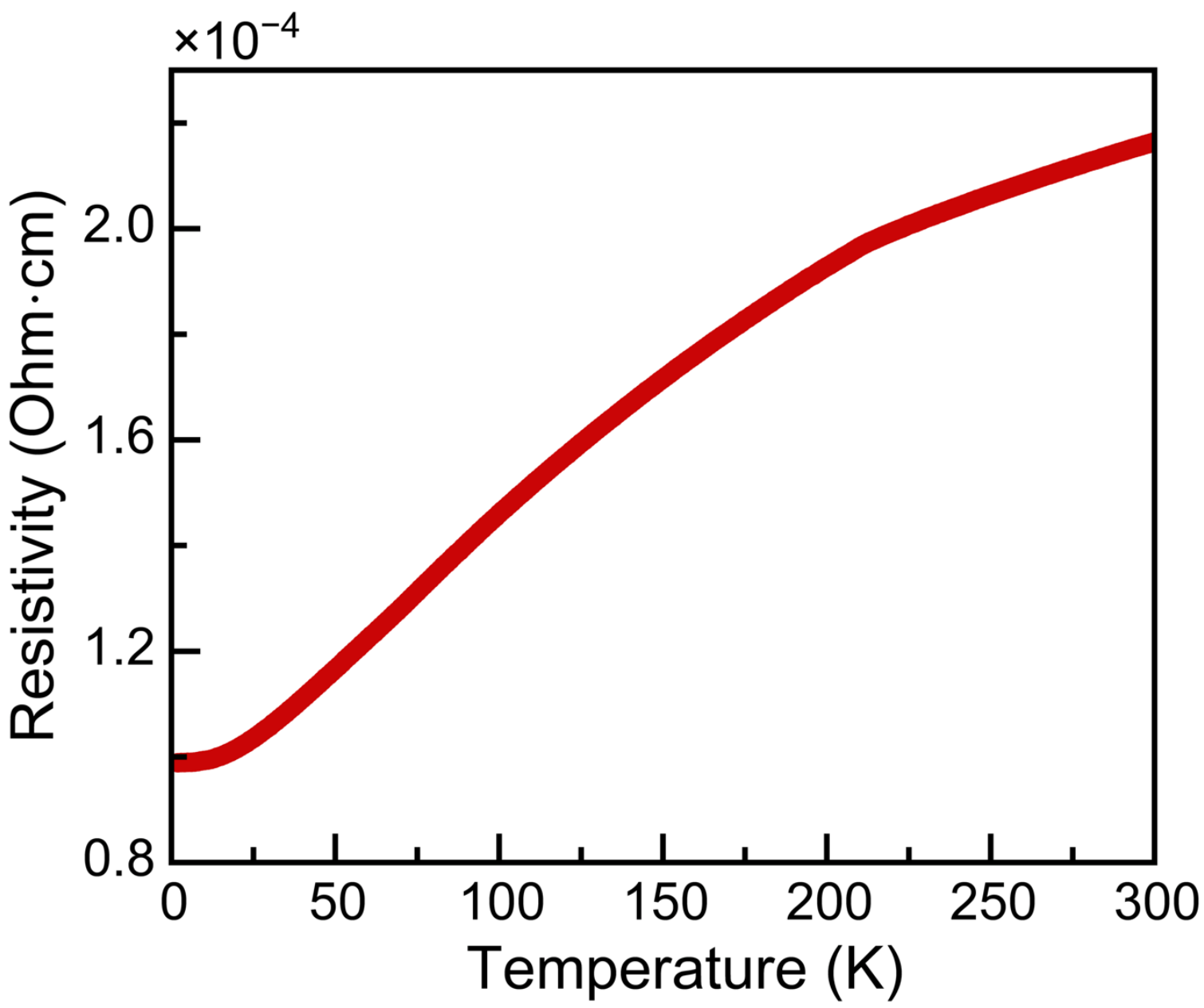

**Figure S11.** Resistivity versus temperature of a $Cr_7Se_8$ single-crystal piece.

**Table S1. Crystal structure refinement data and parameters from single-crystal X-ray diffraction (SXRD)**

| Space group | $P6_3/mmc$ (No. 194) |
|---|---|
| Crystal system | hexagonal |
| Formula (refined) | $Cr_{6.896\pm0.096}Se_8$ |
| $a$ (Å) | 3.6720(2) |
| $c$ (Å) | 5.9936(4) |
| $V$ (Å$^3$) | 69.986 (10) |
| $Z$ | 2 |
| Temperature | 100(2) K |
| Crystal Size (um) | 50, 40, 20 |
| $\lambda$ | 0.71073 |
| $\mu$ (mm$^{-1}$) | 32.485 |
| Absorption Correction | Gaussian |
| $T_{min}$, $T_{max}$ | 0.293, 0.563 |
| Reflections measured, unique, used | 468, 47, 46 |
| Resolution ($d_{min}$, $2\theta_{max}$) | 0.427, 56.248 |
| $(hkl)_{max}$ | (–4, 4), (–4, 4), (–7, 7) |
| Parameters | 6 |
| Method of refinement | Shelx, $\|F^2\|$ |
| $wR_2$ (all) | 0.0545 |
| $R_1$ (all) | 0.0228 |
| GOF | 1.429 |
| $F_{000}$ | 110 |
| Max residual density | 0.590, –1.048 (REM Highest difference peak) |
| CSD code | TBD |

**Table S2. Atomic site coordinates and parameters from SXRD**

| Atom | Site | *x* | *y* | z | $U_{iso}$ (Å$^2$) | *Occ.* |
|---|---|---|---|---|---|---|
| Cr1 | Cr | 0 | 0 | 0.5 | 0.0083(9) | 0.862(12) |
| Se2 | Se | 0.66667 | 0.33333 | 0.75 | 0.0067(5) | 1 |

**Table S3. Anisotropic thermal parameters from SXRD**

| Atom | $U_{11}$ | $U_{22}$ | $U_{33}$ |
|---|---|---|---|
| Cr1 | 0.0096(11) | 0.0096(11) | 0.0058(16) |
| Se1 | 0.0041(6) | 0.0041(6) | 0.0117(8) |

## Supplementary Reference